\documentclass[aps,pre,twocolumn,superscriptaddress,showpacs,nofootinbib,floatfix]{revtex4}

\usepackage{dcolumn}
\usepackage{amsmath}
\usepackage{amssymb}
\usepackage{amsthm}

\usepackage{graphicx}
\usepackage{bm}
\usepackage[T1]{fontenc}
\usepackage{color}
\usepackage{url}
\usepackage[bf]{subfigure}
\usepackage{rotating}
\usepackage{scalefnt}
\usepackage{multirow}
\usepackage{scalefnt}

\newcommand{\lrangle}[1]{\langle{#1}\rangle}

\newcommand{\av}[1]{\langle {#1} \rangle}

\newcommand{\Ct}{C_{\mathrm{th}}}

\begin{document}

\title{An overview of epidemic models with phase transitions to absorbing states running on top of complex networks}

\author{Ang\'{e}lica S. Mata }
\email{angelica.mata@ufla.br}
\affiliation{Departamento de F\'{\i}sica, Universidade Federal de Lavras, Caixa postal 3037, CEP:37200-900, Lavras, Minas Gerais, Brazil}

\date{\today}

\vspace{1cm}

    \begin{abstract}
Dynamical systems running on the top of complex networks has been extensively investigated for decades. But this topic
still remains among the most relevant issues in complex network theory due to its range of applicability.
The contact process (CP) and the susceptible-infected-susceptible (SIS) model are used quite often
to describe epidemic dynamics. Despite their simplicity, these models are robust to predict the kernel of real situations. 
In this work, we review concisely both processes that are well-known and very applied examples of models that exhibit 
absorbing-state phase transitions. In the epidemic scenario, individuals can be infected or susceptible. A
phase transition between a disease-free (absorbing) state and an active stationary phase (where a fraction of the population 
is infected) are separated by an epidemic threshold. For the SIS model, the central issue is to determine this
epidemic threshold on heterogeneous networks. For the CP model, the main interest is to relate critical
exponents with statistical properties of the network.

\end{abstract}
~\\

  \maketitle
\section{Introduction}
\label{intro}

The present paper briefly reviews the modeling and theory of non-equilibrium dynamical systems on 
networks. A key class of non-equilibrium process are those that exhibit
absorbing states, {\it i.e.} states from which the dynamics cannot escape once
it has fallen onto them. 
A relevant feature of
systems that presents absorbing states is a non-equilibrium
phase transitions among an active state, in which the activity lasts forever 
in the thermodynamic limit, and
an absorbing state, in which activity is absent \cite{henkel08,Marrobook}. 

The same type of 
transition occurs in epidemic spreading processes~\cite{epidemics} since a fully healthy state
is absorbing in the above sense, provided that
spontaneous birth of infected individuals is not allowed. The susceptible-infected-susceptible (SIS)~\cite{anderson92}
model and the contact process (CP)~\cite{harris74} represent some of the simplest epidemic models possessing an
absorbing-state phase transition. 

Lattice systems that exhibit such absorbing state phase transitions have universal features, determined 
by conservation laws and symmetries which allow to group them in a same universality 
class\footnote{Once a set of models share the same symmetry properties, irrespective of the microscopic
details of their dynamical rules, they belong to a single universality class and they should have the same critical 
exponents and scaling functions.}~\cite{henkel2008non}. The most robust class of absorbing state phase transitions 
is the directed percolation
that was originally introduced as a model for directed random connectivity~\cite{ref87henkel2008non}. Both 
CP and SIS models are interacting particle systems involving self-annihilation and catalytic creation 
of particles that presents an absorbing-phase transition and thus belong to directed percolation class. The SIS dynamics is indeed the 
most studied model to describe epidemic spreading on networks. Although the CP was initially thought as a toy model 
for epidemics, lately it has been widely used as a generic reaction-diffusion model 
to study phase transition with absorbing states.

Other epidemic model that also presents an absorbing phase transition is the susceptible-infected-recovered-susceptible 
model (SIRS)~\cite{anderson92}.
It is an extension of the standard SIS model, allowing a temporary immunity of nodes. Both SIS and SIRS models are 
equivalent from the mean-field theory perspective, but the mechanism of immunization changes the behavior of the epidemic dynamics 
depending on the heterogeneity of the network structure. The susceptible-infected-recovered (SIR) model is another example of epidemic models 
with permanent immunity, it means that a recovered node can no longer return to the susceptible compartment, so the system present many
absorbing states since each configuration that have only susceptible and recovered nodes is 
absorbing~\cite{Dorogovtsev08, Pastor01}.

In the face of this context, we reviewed the SIS and CP models 
as examples to investigate absorbing phase transitions in 
complex networks. We firstly explain, in section \ref{sec:networks}, some basic concepts related to complex networks 
required to understand the main idea of this paper. Then we describe both epidemic models in section \ref{sec:epidemicmodels} and, in the 
section \ref{sec:meanfield}, we present distinct theoretical approaches 
devised for them. In 
section \ref{sec:SIS_simulation}, we described some commonly used simulation techniques to analyze both models numerically. For the SIS model, 
the central issue is to determine an epidemic threshold separating an absorbing, disease-free state from an active
phase on heterogeneous networks~\cite{Castellano10,Goltsev12,Ferreira12,Gleeson11,Lee2013,Cator12a,mata2013pair,odor2013spectral,boguna2013nature}.
While for the CP model, most of the interest is to relate the critical exponents with statistical properties of the network, 
in particular the degree distribution \cite{Castellano08,Hong2007,Boguna09,cpannealed,FFCR11,Castellano:2006}. In sections
\ref{sec:HMF_CP_openproblem} and \ref{sec:QMFversusHMF} we present a discuss about these points 
related to CP and SIS models, respectively. Finally, in section 
\ref{sec:conclusions} we draw our final comments.

\section{Complex Networks}
\label{sec:networks}

Network analysis is a powerful tool that provide us a fruitful framework to describe phenomena related to
real-world complex systems. Here we will describe just some features of complex networks that it will be 
used throughout the paper. We will also present the uncorrelated configuration model (UCM)~\cite{Catanzaro05}, 
the substrate that will be used to model the dynamics of the epidemic process on networks.

\subsection*{Basic Concepts}

 We can represent a network by means of an adjacency matrix ${\bf A}$. 
A graph of $N$ vertices has a $N \times N$ adjacency matrix. The edges can be represented by the elements $A_{ij}$ 
of this matrix such that~\cite{caldarelli2007sfn}

\begin{equation}
\label{eq:adjacencymatrix}
A_{ij} = \left\{\begin{array}{lc}
1, ~~~ \text{if the vertices $i$ and $j$ are connected}\\
0, ~~~~~~  \text{otherwise}, 
\end{array}\right.
\end{equation}
for a undirected and unweighted graph. In this case, the adjacency matrix is symmetric, it means $A_{ij} = A_{ji}$.

A relevant information gives from the adjacency matrix is the degree $k_i$ of a vertex $i$ defined as the 
number of links that the vertex $i$ has, {\it i.e.}, the number of nearest neighbors of the vertex $i$. The degree of the 
vertices can be written as~\cite{barratbook}

\begin{equation}
k_{i} = \sum_{j=1}^{N} A_{ij}.
\end{equation} 

When it concerns to very large systems a suitable description can be done by means of statistical measures as
the degree distribution $P(k)$. 
The degree distribution provides the probability that a vertex chosen at random has $k$ edges~\cite{Albert02,Dorogovtsev:2002}.
The average degree is an information that can be extracted from $P(k)$ and it is given by the average value of $k$ 
over the network, it means,

\begin{equation}
\langle k \rangle = \frac{1}{N} \sum_{i=1}^{N} k_i = \sum_k k P(k).
\label{eq:averagedegree}
\end{equation}

Similarly, it can be useful to generalize and calculate the {\it n-th} moment of the degree distribution \cite{barratbook} 

\begin{equation}
 \langle k^{n} \rangle = \sum_{k} k^{n} P(k).
\end{equation}

We can classify networks according to their degree distribution. The basic classes are homogeneous and heterogeneous networks.
The first ones exhibit a fast decaying tail, as for example, a Poisson distribution. Here the average degree value corresponds 
to the typical value in the system. Heterogeneous networks exhibit heavy tail that can be approximated by
a power-law decay, $P(k) \sim k^{-\gamma}$. In this kind of network, the vertices often 
have a small degree, but there is a non-negligible probability of finding nodes with very large degree values thus, 
depending on $\gamma$, the average degree does not represent any characteristic value of the distribution~\cite{barratbook}. 

Many network model have been created in order to describe real systems. The advantage of using 
a model is to reduce the complexity of the real world to a level that one can be treated, for example, 
from the perspective of mean-field approach. In this context, uncorrelated random graphs are important from a
numerical point of view, since we can test the behavior of dynamical systems whose theoretical solution is available 
only in the absence of correlations. For this propose, Catanzaro and collaborators~\cite{Catanzaro05} presented an 
algorithm to generate uncorrelated random networks with  power law degree distributions, called
uncorrelated configuration model (UCM) as described below. 

\subsection*{Uncorrelated Configuration Model}

To construct this networks we started with a set of $N$ disconnected vertices. Each node $i$ is signed with a number 
$k_{i}$ of stubs, where $k_i$ is a random  variable with distribution $P(k) \sim k^{-\gamma}$ under the restrictions 
$k_0 \leq k_i \leq N^{1/2}$ and $\sum_i k_i$  even. It means that no vertex can have either a degree smaller than the 
minimum degree $k_0$ or larger than the cutoff $k_c = N^{1/2}$. The network is constructed by randomly choosing two stubs 
and connecting them to form links, avoiding both self and multiple connections~\cite{Catanzaro05}.

It is possible to show that~\cite{mariancutofss}, to avoid correlations in the absence of multiples and self-connections, 
the random network must have a {\it structural cutoff} scaling at most as $k_c(N) \sim N^{1/2}$. As said previously, 
this algorithm is very useful in order to check the accuracy of many analytical solutions of dynamical process on networks. 
Because of that, it was chosen as a substrate for implementing the dynamics of the SIS and CP models in this review work.

\label{sec:networks}

 \section{Epidemic Models}
\label{sec:epidemicmodels}

In the SIS epidemic model, each vertex $i$ of the network can be only one of two states: infected or susceptible. 
Let us assume the most general case where a vertex $i$
becomes spontaneously healthy at rate $\mu_i$, and transmits the infection to each one of its $k_i$ neighbors at rate $\lambda_i$.
For classical SIS, one has $\mu_i = \mu$ and 
$\lambda_i = \lambda$ for every vertex~\cite{Pastor01}.

 
As in the SIS model, vertices in the CP model can
be infected or susceptible, which in reaction-diffusion system's jargon are called occupied and 
empty, respectively.
The spontaneous cure process is exactly the
same as in the SIS model: infected vertices become susceptible at
rate $\mu_i = \mu$. However, the infection is different. An infected
vertex tries to transmit the infection to a randomly chosen
neighbor at rate $\lambda$, implying that the transmission rate of vertex $i$ is $\lambda_i = \lambda/k_i$, where $k_i$ 
is the number of neighbors of the i-th node. This reduces drastically the infective power of very connected vertices in comparison with the SIS dynamics.

\begin{figure}[ht]
\centering
\includegraphics[width=7.0cm]{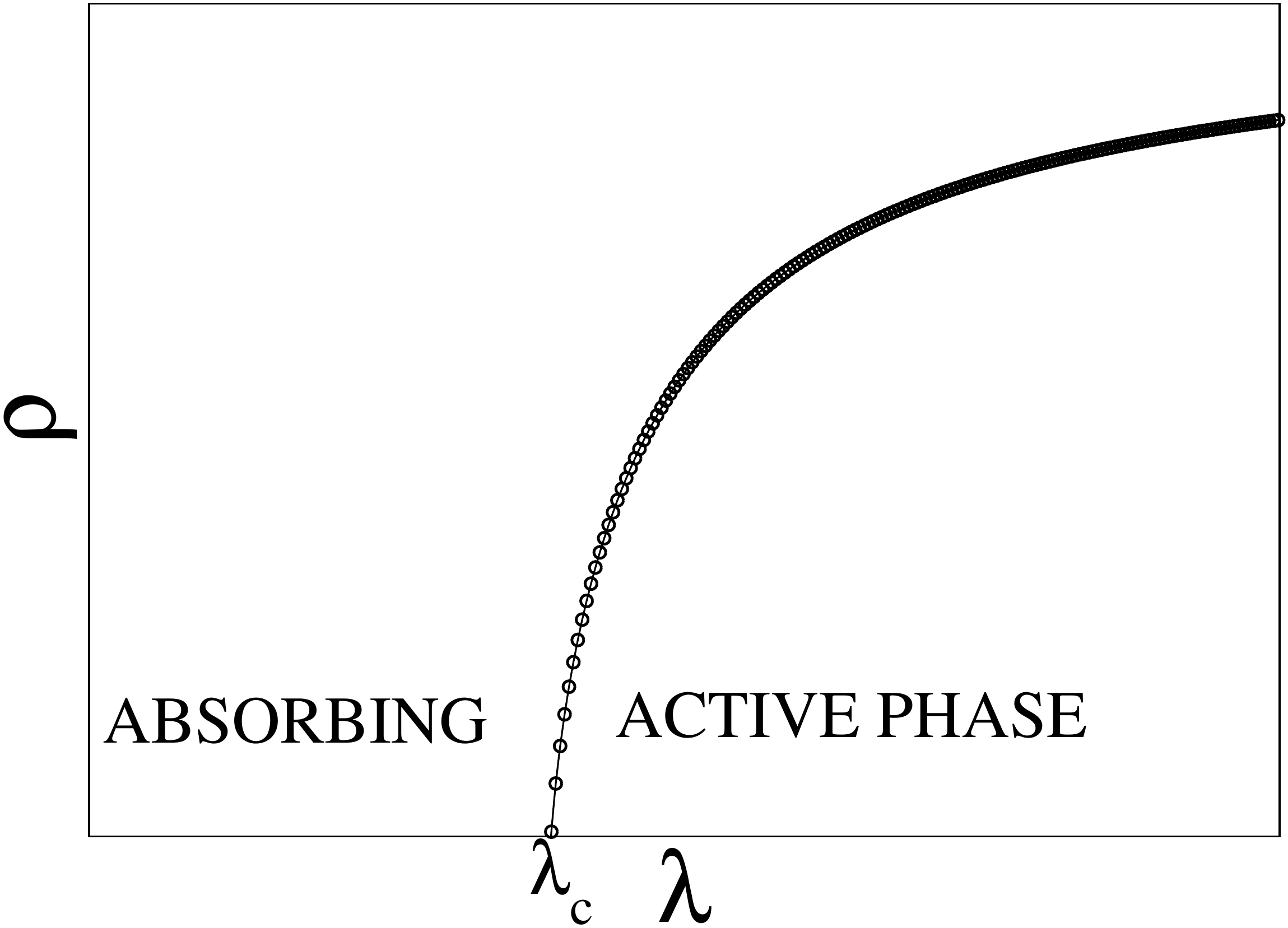}
\caption{The usual behavior of the density of infected nodes $\rho$ in function of the control parameter $\lambda$,
in a epidemic model as SIS or CP in the thermodynamic limit. The value $\lambda_c$ is the epidemic threshold that
separates an absorbing state to an active phase with $\rho >0$.}
\label{fig:rhoXlambda}
\end{figure}

Both SIS and CP dynamics exhibit a phase transition
between a disease-free (absorbing) state and an active stationary phase where a fraction of
the population is infected. Originally, these regions are separated by an epidemic threshold
$\lambda_c$~\cite{Dorogovtsev08, Pastor01}.
The density of infected nodes $\rho$ is the standard order parameter that describes this phase transition, as shown
in Figure \ref{fig:rhoXlambda}. 
However, for a finite system the unique true stationary state is the absorbing state, even above the critical point, due to dynamical fluctuations. To overcome the difficulty to study the active
state of finite systems some simulation strategies were proposed in the literature~\cite{Marrobook,DickmanJPA,DeOliveira05,boguna2013nature}, as we will show in 
section~\ref{sec:absorbing}. 

When we study epidemic processes running on the top of heterogeneous networks a more complex behavior can emerge. Indeed, the accurate theoretical understanding of epidemic models running on the top of complex networks rates among
the hottest issues in the physics community~\cite{Goltsev12, Castellano10, odor2013spectral, Lee2013, boguna2013nature,
mata2013pair, Pugliese09, Gomez10, Gomez:nopert, JuhaszCP, Castellano08, Hong2007, multipletransitions, holme_pqmf2016, Castellano2020, mountford2013,
spectralproperties, Chatterjee09}. 
Much effort has been devoted to understand the criticality of the absorbing state phase transitions observed in 
CP~\cite{JuhaszCP, Castellano08, Hong2007, Boguna09, cpannealed} and SIS~\cite{Goltsev12, Castellano10, odor2013spectral, Lee2013,
boguna2013nature, mata2013pair, Gomez:nopert, holme_pqmf2016, multipletransitions, Castellano2020, mountford2013,
spectralproperties, Chatterjee09} models, mainly based on perturbative approaches (first order in $\rho$)
around the transition point~\cite{Goltsev12,
Castellano10, odor2013spectral, mata2013pair, Gomez10, Castellano08}.
In the following we will present the basic mathematical approaches for the epidemic dynamics.

\section{Mean-field theories}
\label{sec:meanfield}

The prediction of disease evolution can be conceptualized within a variety of mathematical approaches.
All theories aim at understanding the properties of epidemics in the equilibrium or long
term steady state, the existence of a non-zero density of infected individuals, the
presence or absence of a threshold, etc. We will start with the simplest case namely the homogeneous mean-field theory,
and thereafter we will review other more sophisticated mathematical approaches.

\subsection{Homogeneous mean-field theory}

The simplest theory of epidemic spreading assumes that the population can be divided into different compartments 
according to the 
stage of the disease (for example, susceptible and infected in both SIS and CP models) and within each compartment, individuals 
(vertices in the complex networks' jargon) are assumed to be identical and have approximately the same number of 
neighbors (edges), $k \approx \langle k \rangle$. The idea is to write a time evolution equation for the number of
infected individuals $I(t)$, or equivalently the corresponding density $\rho(t) = I(t)/N$, where $N$ is the total number of 
individuals. For example, the equation describing the evolution of the SIS model is~\cite{baileybook}:
\begin{equation}\label{eq:sishomogeneous}
 \frac{d\rho(t)}{dt} = -\mu \rho(t) + \lambda \langle k \rangle \rho(t) [1-\rho(t)],
\end{equation}
considering uniform infection and cure rates ($\lambda_i = \lambda$ and $\mu_i = \mu$, $\forall ~~ i = 1, 2, \cdots N$). 
The first term on the right-hand side of the Eq.~\eqref{eq:sishomogeneous} refers to the spontaneous healing and the second one to the infection process, that is proportional to the spreading 
rate $\lambda\langle k \rangle$, the density of the susceptible vertices is $1-\rho(t)$ that may become infected, and the
density of infected nodes $\rho(t)$ in contact with any susceptible individual. 
Note that
the evolution of the SIS model is completely described by Eq.~\eqref{eq:sishomogeneous}, since the density of susceptible 
individuals is $S(t)/N = 1-\rho(t)$. 
The mean-field character of this equation comes from the fact that the correlations 
among different nodes were neglected. Thus, the probability that one infected vertex is connected to a susceptible one is
approached as $\rho(t)[1-\rho(t)]$.

Near the phase transition
between an absorbing state and an active stationary phase we can assume that the number of infected nodes is
small $\rho(t) \ll 1$. In this regime, we can use a linear approximation neglecting all $\rho^2$ 
terms\footnote{Indeed, this equation 
can be exactly integrated, but we prefer this stability analysis that is used in more complex theories in the next subsections.}. So the 
Eq.~\eqref{eq:sishomogeneous} becomes\footnote{From now on, we will consider $\mu = 1 $, unless otherwise specified.}:
\begin{equation}\label{eq:sishomogeneousapprox}
 \frac{d\rho(t)}{dt} = -\rho(t) + \lambda \langle k \rangle \rho(t).
\end{equation}
The solution is $\rho(t) \sim e^{-(1 - \lambda \langle k \rangle)t}$, implying that $\overline{\rho} = 0$ is an stable fixed point for 
$1 - \lambda \langle k \rangle > 0$. Thus, one obtains,
\begin{equation}
 \lambda_c = \frac{1}{\langle k \rangle}.
\end{equation}
Here, $\lambda_c$ is the epidemic threshold such that for any infection rate above this value the epidemic lasts forever~\cite{baileybook}.

Similar analysis can be done for the CP model. In this case, the homogeneous mean-field equation read as
\begin{equation}
  \frac{d\rho(t)}{dt} = -\rho(t) + \lambda\rho(t) [1-\rho(t)],
\end{equation}
since the transmission rate of each node is $\lambda/\langle k \rangle$. Performing the same linear stability analysis in the 
steady-state, one obtains $\lambda_c = 1$.

In this framework, one considers that the connectivity patterns among individuals are homogeneous,
neglecting the highly
heterogeneous structure of the contact network inherent to real
systems~\cite{Albert02}. Many biological, social and technological systems are
characterized by heavy tailed distributions of the number of contacts $k$ of an
individual (the vertex degree), characterized by a power law degree distribution, $P(k) \sim k^{-\gamma}$. For such systems the homogeneity 
hypothesis is severely violated~\cite{Albert02,Newman10,barratbook}.  Complex
networks are, in fact, a framework where the heterogeneity of the contacts can be
naturally afforded~\cite{Albert02}. Indeed, this heterogeneity plays the main role in 
determining the epidemic threshold. To take it into account other approaches have been proposed, as we showed in the next
sections. The major aim is to understand how the epidemic spreading can be strongly influenced by the topology of networks.


\subsection{Quenched Mean-field Theory}
\label{sec:QMF}

An important and frequently used mean-field  approach to describe epidemic dynamics on heterogeneous networks is the quenched mean-field (QMF) theory~\cite{Castellano10}, that explicitly takes into account the actual connectivity of the network through
its adjacency matrix. The central idea is to write the evolution equation for the probability $\rho_i(t)$  that a certain 
node $i$ is infected. For the SIS model the dynamical equation for this probability takes the form~\cite{Castellano10}:
\begin{equation}
 \frac{d\rho_i}{dt} = -\rho_i+\lambda(1-\rho_i)\sum_{j=1}^NA_{ij}\rho_j.
\label{eq:rho_i1}
\end{equation}
where $A_{ij}$ is the adjacency matrix.
The first term on the right-hand side 
considers nodes becoming healthy spontaneously while the second one 
considers the event
that the node $i$ is healthy and gets the infection via a neighbor node.

Performing a linear stability analysis around the trivial fixed
point $\rho_i=0$, one has  
\begin{equation}
\frac{d\rho_i}{dt} = \sum_j L_{ij}\rho_j,
\end{equation}
where the Jacobian matrix is
\begin{equation}
L_{ij}=-\delta_{ij}+\lambda A_{ij}, 
\end{equation}
$\delta_{ij}$ being the Kronecker delta symbol. 
The transition occurs when the fixed point loses stability or,
equivalently, when the largest eigenvalue of the Jacobian matrix is 
$\Upsilon_m=0$~\cite{Hilborn}. The largest eigenvalue of $L_{ij}$  is given by
$\Upsilon_m=-1+\lambda \Lambda_m$ where $\Lambda_m$ is the largest eigenvalue of
$A_{ij}$.  Since $A_{ij}$
is a real non-negative symmetric matrix, the Perron-Frobenius theorem states that 
one of its eigenvalues is positive and greater than, in absolute value, all other eigenvalues, 
and its corresponding eigenvector has positive components. 
So, one obtains the epidemic threshold of the SIS model in a QMF approach~\cite{Castellano10}:          
\begin{equation}
\lambda_c^{qmf}={1}/{\Lambda_m},
 \label{eq:lbcqmf}
\end{equation}
where $\Lambda_m$ is the largest eigenvalue of the adjacency matrix.
For the CP dynamic, the Eq.~\eqref{eq:rho_i1} becomes,
\begin{equation}
 \frac{d\rho_i}{dt} = -\rho_i+\lambda(1-\rho_i)\sum_{j}\frac{A_{ij}\rho_j}{k_j}. 
\end{equation}
Performing the same linear stability analysis around the trivial fixed
point, as was done for the SIS model, one obtains
\begin{equation}
 \frac{d\rho_i}{dt} = \sum_j L_{ij}\rho_j,
 \label{eq:rho_iOneLin}
\end{equation}
where the Jacobian matrix is given by
\begin{equation}
L_{ij}=-\delta_{ij}+\frac{\lambda A_{ij}}{k_j}.
 \label{eq:Jaco1}
\end{equation}
Once again the transition point is defined when the absorbing 
phase becomes unstable or, equivalently, when the largest eigenvalue  
of $L_{ij}$ is null~\cite{Hilborn}. The largest eigenvalue of $L_{ij}$  is given by
$\Upsilon_m=-1+\lambda \Lambda_m$ where $\Lambda_m$ is the largest eigenvalue of
$C_{ij}=A_{ij}/k_j$. Notice that $v_i=k_i$ is an eigenvector of $C_{ij}$ with
eigenvalue $\Lambda=1$. Now, supported by the Perron-Frobenius theorem~\cite{Newman10},
we conclude that the largest eigenvalue of $C_{ij}$ is $\Lambda_m=1$ resulting
in the transition point $\lambda_c=1$, as obtained in a homogeneous approximation.

Returning to the SIS model, the equation \eqref{eq:lbcqmf} can be complemented with the results of Chung {\it et. al.}~\cite{chung03} 
who calculated the largest
eigenvalue of adjacency matrix of 
networks with a power law degree distributions  as
\begin{equation}
\Lambda_m \simeq \left\{\begin{array}{lc}
\sqrt{k_{c}}, ~~~ \gamma > 5/2\\
\frac{\langle k^2 \rangle}{\langle k \rangle}, ~~~ 2 < \gamma < 5/2
\end{array}\right.
 \label{eq:lbc_qmf}
\end{equation}
where $k_c$ is the degree of the most connected node and $\langle k^n \rangle$ is the {\it n-th} moment of the degree distribution. Since $k_c$ grows as a function of the network size for any $\gamma$ and $\langle k^2 \rangle$ diverges for $2< \gamma <3$, the central result of equation~\eqref{eq:lbc_qmf} is: $\Lambda_m$ diverges for
enlarging networks with power law degree distributions
even when $\lrangle{k^2}$ remains finite~\cite{chung03}. Therefore, the epidemic thresholds scale as \cite{Castellano12}
\begin{equation}
\lambda_c \simeq \left\{\begin{array}{lc}
1/\sqrt{k_{c}}, ~~~ \gamma > 5/2\\
\frac{\langle k \rangle}{\langle k^2 \rangle}, ~~~ 2 < \gamma < 5/2
\end{array}\right.
 \label{eq:lbc_qmf2}
\end{equation}
which vanishes for any power-law degree distribution. The reasons for this difference of $\lambda_c$ predicts by QMF 
approach, for $\gamma$ larger or smaller than $5/2$ are explained in 
ref.~\cite{Castellano12}. In processes allowing endemic steady-states, the activation mechanisms depend on the degree of heterogeneity 
of the network. For $\gamma>5/2$ the hub sustains activity and propagates it to the rest of the system while for $\gamma<5/2$ the 
innermost network core collectively turns into the active state maintaining it globally~\cite{Castellano12, PhysRevXRomu, phasetransitionSIS2018}.  
However, the behavior of the SIS model 
on random networks with power-law degree distribution can be much more complex than previously 
thought. We will discuss these different possible scenarios in section \ref{sec:QMFversusHMF}.

Although mean-field theories are a simplified description of models, it is expected that they
correctly predicts the behavior of dynamical process on networks, due to its small-world
property. However, dynamical correlations are not taken into account since the states of a node and its 
neighbors are considered independent. One can consider dynamical correlations by means of a
pair-approximation~\cite{Marrobook} in which the dynamic of an individual is explicitly
influenced by its nearest neighbors as we showed in the section \ref{sec:pairqmf}.

\subsection{Heterogeneous Mean-Field Theory}
\label{sec:HMF}


In the degree-based theories, called heterogeneous mean-field (HMF) theory,  dynamical
quantities, as the density of infected individuals, depend only
of the vertex degree and do not of their specific
location in the network. Actually, the HMF theory can be obtained from the QMF one performing a coarse-graining 
where vertices are grouped according to their degrees. To 
take into account the effect of the degree heterogeneity we have to consider
the relative density $\rho_k(t)$ of infected nodes with a given degree $k$, {\it i.e.}, the probability that a node
with $k$ links is infected. Again using the SIS model as an example, the dynamical mean-field rate 
equation describing the system can thus be written as \cite{Pastor01}:
\begin{equation}
\frac{d\rho_k(t)}{dt} = -\rho_k(t)+\lambda k[1-\rho_k(t)]\sum_{k'}P(k'|k) \rho_{k'}(t),
 \label{eq:rhok1_cap2}
\end{equation}
The first term on the right-hand side considers nodes becoming healthy at unitary rate. The second term considers the event
that a node with $k$ links is healthy and gets the infection via a nearest neighbor. The probability of this event is proportional
to the infection rate $\lambda$, the number of connections $k$ and the probability that any neighbor vertex is 
infected $P(k'|k) \rho_{k'}$. The linearization of Eq.~\eqref{eq:rhok1_cap2} gives
\begin{equation}
 \frac{d\rho_k}{dt} = \sum_{k} L_{k k'} \rho_{k'},
\end{equation}
where $L_{k k'} = -\delta_{k k'} + \lambda k P(k'|k)$. Therefore, the epidemic threshold is 
\begin{equation}
 \lambda_c = \frac{1}{\Lambda_m},
\end{equation}
where $\Lambda_m$ is the largest value of $C_{k k'} = k P(k'|k)$.

It is difficult to find the exact solution for $\Lambda_m$ for a general form of $P(k'|k)$. 
But it is possible to extract the value of the epidemic threshold.  In the case of uncorrelated networks, 
$P(k'|k) = k'P(k')/\langle k \rangle$ and $C_{k k'} = k' k P(k')/\langle k \rangle$. So, it is easy to check that $v_k = k$ 
is an eigenvector with eigenvalue  $\langle k^2 \rangle/ \langle k \rangle$ that, according to Perron-Frobenius theorem, is the 
largest. Thus, we obtain the epidemic threshold: 

\begin{equation}
\lambda_c^{hmf}={\lrangle{k}}/{\lrangle{k^2}},
 \label{eq:lbchmf}
\end{equation} 

Equation~(\ref{eq:lbchmf}) has strong
implications since several real networks have a power law degree distribution
$P(k)\sim k^{-\gamma}$ with exponents in the range
$2<\gamma<3$~\cite{Albert02}. For these distributions, the second moment $\lrangle{k^2}$ diverges in the
limit of infinite sizes implying a  vanishing threshold for the
SIS model or, equivalently,  the epidemic prevalence for any finite infection
rate. 
Both theories HMF and QMF predict vanishing thresholds for $\gamma<3$ despite of different
scaling for $5/2<\gamma<3$. However, HMF predicts a finite threshold for
networks with $\gamma>3$ unlike the QMF theory that still predicts a
vanishing threshold~\cite{Castellano12}. 

For the CP model, the heterogeneous mean-field equation, analogous to Eq.~\eqref{eq:rhok1_cap2}, can be written as
\begin{equation}
\frac{d\rho_k(t)}{dt} = -\rho_k(t)+\lambda k[1-\rho_k(t)]\sum_{k'}\frac{P(k'|k) \rho_{k'}(t)}{k'}.
 \label{eq:rhok1_cap2_CP}
\end{equation}
Again we assume degree-uncorrelated networks and a simple linear 
stability analysis shows the presence of a phase transition, located at the
value $\lambda_c = 1$, as found in homogeneous mean-field theory, independent of the degree 
distribution and
degree correlations. 
According to simulations~\cite{Boguna09, JuhaszCP,Castellano08,Castellano:2006, FFCR11,cpannealed} the transition point does not quantitatively reproduced the predictions of
both approaches, HMF and QMF.
However, the advantage of HMF over the QMF theory, is that we can analytically obtain the critical exponents for the dynamical 
model and compare with numerical results.
Indeed, some works have shown that the contact process running on the top of highly
heterogeneous random networks is well-described by the heterogeneous mean-field
theory~\cite{FFCR11,cpannealed}. However, some important aspects such as the threshold and strong
corrections to the finite-size scaling observed in simulations are not clarified in this theory.
We summarized the intense scientific discussion~\cite{Hong2007, Castellano08,Castellano:2006,
comment1, comment2,FFCR11,cpannealed} about this subject in section \ref{sec:HMF_CP_openproblem}.

 \subsection{Pair Mean-Field Theories}

 Improvement of both HMF and QMF theories including dynamical correlations by means
of a pair-approximation do not change qualitatively the results, however they 
promote a quantitative refinement, as we showed hereinafter.

 \subsubsection{Pair Quenched Mean-Field Theory}
\label{sec:pairqmf}

In the paper \cite{mata2013pair}, the authors investigated the role of 
dynamical correlations on the dynamic of the SIS epidemic model on different substrates.
We can start rewriting Eq.~\ref{eq:rho_i1} as follows:

\begin{equation}
 \frac{d\rho_i}{dt} = -\rho_i+\lambda\sum_j\phi_{ij}A_{ij},
\label{eq:rho_iPairQMF}
\end{equation}

where $A_{ij}$ is the adjacency matrix and $\phi_{ij}$ represents the probability of a 
pair of nodes $i$ and $j$ of being in the state  $\phi_{ij} = [S_iI_j]$, this means the node 
$i$ is susceptible and its neighbor $j$ is infected. When we uses a simple approximation, 
this joint probability was factorized: $\phi_{ij}\approx (1-\rho_i)\rho_j $. Then, we 
should write a dynamical equation for $\phi_{ij}$:

\begin{equation}
 \frac{d\phi_{ij}}{dt}=-\phi_{ij}-\lambda\phi_{ij}+[I_iI_j] \nonumber
\end{equation}
 \begin{equation}
+\lambda\sum_{l \ne i}             
 [S_i,S_j,I_l]A_{jl}
-\lambda\sum_{l\ne j}
[I_l,S_i,I_j]A_{il}.
\label{eq:phi1QMF}
\end{equation}

The first three terms represents the processes related to the pair of neighbors $i$ and $j$,
spontaneous annihilation reactions $[S_i,I_j]\rightarrow[S_i,S_j]$ 
and $[I_i,I_j]\rightarrow[S_i,I_j]$ and the infection in vertex $i$ due to 
$j$,  $[S_i,I_j]\rightarrow[S_i,S_j]$. The other terms represent 
processes related to the interaction with the other neighbors of $i$ and $j$, this means
another vertex $l$ that can infect $i$ or $j$: $[I_l,S_i,I_j]\rightarrow [I_l,I_i,I_j]$ and 
$[S_i,S_j,I_l]\rightarrow [S_i,I_j,I_l]$. 
Equations~(\ref{eq:rho_iPairQMF}) and (\ref{eq:phi1QMF}) cannot be solved due to the
triplets. In turn, the dynamical equations for triplets will depend on
quadruplets, and so on. Then we have to break these correlations in some point. To obtain an
pair-approximation solution, we can apply the standard 
pair-approximation~\cite{Avraham92,henkel08}:
\begin{equation}
 [A_i,B_j,C_l]\approx \frac{[A_i,B_j][B_j,C_l]}{[B_j]}.
\label{eq:cluQMF}
\end{equation}

After a few steps (details are in reference~\cite{mata2013pair}), similar to what we did in the one-vertex mean-field approximaation as to perform a linear stability 
analysis around the fixed point $\rho_i = [S_iI_j] = [I_iI_j] = 0$ and a quasi-static approximation for 
for $t\rightarrow\infty$, $d\rho_i/dt\approx 0$ and $d\phi_{ij}/dt\approx 0$, in Eqs.~(\ref{eq:rho_iPairQMF}) and
(\ref{eq:phi1QMF}), we can find the Jacobian matrix
\begin{equation}
 L_{ij} = -\left(1+\frac{\lambda^2 k_i}{2\lambda+2}\right)\delta_{ij}
+\frac{\lambda(2+\lambda)}{2\lambda+2}A_{ij}.
\label{eq:Lij}
\end{equation}
As in the case of one-vertex QMF theory, the critical point is obtained when the largest eigenvalue of
$L_{ij}$ is nul. Analytical solution for simple networks as random regular networks, star an
wheel graphs can be directly obtained from Eq.~\ref{eq:Lij}. For power law arbitrary random networks, 
as UCM model, the largest eigenvalue of Eq.~\ref{eq:Lij} can be numerically determined. 
In reference~\cite{mata2013pair}, Mata and Ferreira showed that the thresholds obtained in pair and one-vertex 
QMF theories have the same scaling with the system size but the pair QMF theory is quantitatively 
much more accurate than the one-vertex theory when compared with simulations. In reference~\cite{holme_pqmf2016}
the authors also studied the impact of dynamic correlations on the SIS dynamics on static networks.

Performing the same analysis for the contac process, one obtains the Jacobian matrix:
\begin{equation}
L_{ij}=-(1+\lambda^2\alpha_i)\delta_{ij}+\frac{\lambda(2k_i+\lambda)A_{ij}}{
2k_ik_j+\lambda(k_i+k_j)}. 
\label{eq:Jaco2}
\end{equation}
The critical point is also obtained when its largest eigenvalue is null.
A general analytical expression is not available for large random power-law degree networks
but, in principle, it can be  obtained numerically for any kind of network.
However, for simple graphs as a random regular network, the transition point can 
be obtained after some algebraic manipulations and it is given by:

\begin{equation}
 \lambda_c=\frac{m}{m-1},
\label{eq:lbc_RR}
\end{equation}
where $m$ is the 
degree of all nodes of the network. That is the same value yield by the simple homogeneous pair-approximation~\cite{Marrobook}.

 \subsubsection{Pair Heterogeneous Mean-Field Theory}
\label{sec:pairhmf}

For the SIS model, the equation for the probability that a vertex with degree $k$ is
occupied takes the form
\begin{equation}
\frac{d\rho_k}{dt} = -\rho_k+\lambda k\sum_{k'}\phi_{kk'} P(k'|k),
 \label{eq:rhok1}
\end{equation}
where the conditional probability $P(k'|k)$, which gives the probability that a
vertex of degree $k$ is connected to a vertex of degree $k'$, weighs the
connectivity between compartments of degrees $k$ and $k'$ and $\phi_{kk'}$ represents
a pair of nodes with degree $k$ and $k'$ respectively, in the state $[S_{k}I_{k'}]$.
 The dynamical equation for $\phi_{kk'}$  is
\begin{equation}
\frac{d\phi_{kk'}}{dt} = -\phi_{kk'}-\lambda \phi_{kk'} k'+ [I_{k}I_{k'}] \nonumber
\end{equation}
\begin{equation}
+ \lambda(k'-1)\sum_{k''} [S_kS_{k'}I_{k''}]P(k''|k')\nonumber
\end{equation}
\begin{equation}
  -\lambda(k-1)\sum_{k''}[I_{k''}S_{k}I_{k'}]P(k''|k).
 \label{eq:phikk1}
\end{equation}

The one-vertex mean-field equation [Eq.~(\ref{eq:rhok1_cap2})] is
obtained factoring the joint probability $\phi_{kk'}\approx (1-\rho_k)\rho_{k'}$
in Eq.~(\ref{eq:rhok1}). The factor $k'-1$ preceding the first
summation in Eq.~(\ref{eq:phikk1}) is due to the $k'$ neighbors of middle
vertex except the link of the pair $[0_k0_{k'}]$ (similarly for $k-1$ preceding
the second summation).

Following the same line of reasoning as the previous calculations, we now approximate 
the triplets in Eq.~(\ref{eq:phikk1}) with the
pair-approximation of Eq.~\eqref{eq:phikk1}, performing a linearization 
 around the fixed point $\rho_k \approx 0$ and $\phi_{kk'} \approx 0$,
and performing a quasi-static approximation for
$t\rightarrow\infty$, in which $d\rho_k/dt \approx 0$ and $d\phi_{kk'}/dt \approx 0$.

Considering uncorrelated random networks, we obtain the Jacobian $L_{kk'}$
\begin{equation}
L_{kk'} = -\delta_{kk'}+  \frac{\lambda k}{\langle k \rangle}\frac{(2k'-1)}{(2+\lambda)}
\label{eq:Jaco2chap2} 
\end{equation}
with $\delta_{kk'}$ being the Kronecker delta symbol.

Again, the absorbing state is unstable when the largest eigenvalue of $L_{kk'}$ is
positive. Therefore, the critical point is obtained when the largest eigenvalue
of the Jacobian matrix is null, thus obtaining:
\begin{equation}\label{eq:HMF_pair_SIS}
\lambda_c = \frac{\lrangle{k}}{\lrangle{k^2}-\lrangle{k}}.
\end{equation}
This threshold coincides with that of the susceptible-infected-recovered (SIR)
model in a one-vertex HMF theory~\cite{barratbook}. This results was also
proposed in Ref.~\cite{boguna2013nature} using heuristic arguments.
They argued that, dynamical correlations, that are neglected in a one-vertex HMF theory, account 
for the fact that infected nodes have higher probability to be still infected. This means that, in the next step,
this node can be considered immunized (recovered for a while). So, a better upper bound for the spreading of
the disease is given by the HMF theory of the SIR model, that is exactly the threshold given by Eq.~\eqref{eq:HMF_pair_SIS}.

In a similar way, we can perform the same analysis for the CP model and we obtain the Jacobian:
\begin{equation}
L_{kk'} = -\delta_{kk'}+  \frac{\lambda k(2k'-1)P(k'|k)}{(2k'+\lambda)k' }
 = -\delta_{kk'}+C_{kk'}.
\label{eq:Jaco2_CP} 
\end{equation}
Assuming that the network does not present correlated degree, we have $P(k'|k) = k'P(k')/\langle k \rangle$.
So, the $u_k=k$ is an eigenvector of $C_{kk'}$ with eigenvalue
\begin{equation}
 \Lambda =
\frac{\lambda}{\lrangle{k}}\sum_{k'}\frac{(2k'-1)P(k')k'}{(2k'+\lambda)}.
 \label{eq:Lambda_CP}
\end{equation}
Since $C_{kk'}>0$ is irreducible and $u_k>0$, the Perron-Frobenius theorem~\cite{Newman10} warranties that $\Lambda$
is the largest eigenvalue of $C_{kk'}$. The critical point is given by $-1+\Lambda=0$, it means we have to solve 
the transcendent equation
\begin{equation}
 \frac{\lambda_c}{\lrangle{k}}\sum_{k'}\frac{(2k'-1)k'P(k')}{(2k'+\lambda_c)}=1,
 \label{eq:lbc_CP}
\end{equation}
numerically to obatin the transition point for any kind of uncorrelated degree network.
Using a random regular network as an example, this means, $P(k) = \delta_{k,m}$, where $m$ is the 
degree of all nodes of the network, we obtain again:
\begin{equation}
 \lambda_c=\frac{m}{m-1},
\label{eq:lbc_hom}
\end{equation}
that is the same of the homogeneous and quenched pair mean-fiel theory.

In reference \cite{Mata14} the authors have also determined the critical exponents 
in the pair HMF approach for the CP model. For the
infinite size limit the exponents are the same as the one-vertex theory, as
expected. However, the finite-size corrections to the scaling obtained in the pair HMF
theory allowed a remarkable agreement with simulations for all degree
exponents ($2.0\le \gamma\le3.5$) investigated, suppressing a deviation observed for low degree exponents in
the one-vertex HMF theory~\cite{FFCR11}.

\section{Simulation of Epidemic Processes}
\label{sec:SIS_simulation}

Numerical simulations is an essential tool to predict the accuracy of mean-field approaches in the study of
epidemic processes on complex networks. Although this tool is widely used, strict implementations of epidemic processes 
on networks with high heterogeneity on degree distribution are not simple. In reference \cite{COTA2018}, the authors showed 
that, depending on the network properties, the threshold of the SIS model can be altered when occur modifications in
the SIS dynamics, even preserving the basic properties of spontaneous healing and 
potencial of infection of each vertex growing unlimitedly with its degree.

The classical algorithm to model continuous-time Markov processes is known as Gillespie algorithm~\cite{Gillespie1976403}. 
In this recipe, we associated each dynamical transition (infection and healing for the SIS or CP model, for example) with a 
Poisson process, this mean, independent spontaneous processes. At each change of state, we have to update a list 
containing all possible spontaneous processes. However, for very large networks, this is computationally unfeasible. So, we used 
an optimized Gillespie algorithm proposed by Cota and Ferreira~\cite{COTA2017}. In the following sections, we summarized 
these optimized algorithms for the SIS and CP dynamics. For details of how to implement 
optimized algorithms of continuous-time Markovian processes based on Gillespie algorithm see reference~\cite{COTA2017}.

\subsection{Simulation of SIS model}

The SIS dynamics in a network of size $N$ can be simulated in a very simple way: Select a vertex at random with equal chance. 
If the selected vertex $i$ is infected we turn it to susceptible with probability 
\begin{equation}
p_i = \frac{\lambda n_i}{(\mu+\lambda)k_{max}},
\end{equation}
where $k_{max}$ is maximal number of connections, 

\noindent $n_i=\sum_jA_{ij}\sigma_j$ is the number of infected nearest neighbors of the vertex $i$, and
$\sigma_j=1$ corresponds to infected node and $\sigma_j = 0$ otherwise. Here, we are 
considering the simplest case of $\lambda_i = \lambda$ and $\mu_i = \mu$ for all vertices.
If the selected vertex $i$ is susceptible, it becomes infected with probability 
\begin{equation}
q_i = \frac{\mu}{(\mu+\lambda)k_{max}}.
\end{equation}

This algorithm is accurate and can used for any generic SIS dynamics. However, if one is interested in
regions close to the threshold where the great majority of the vertices are susceptible, the algorithm is very inefficient since
changes happens only in the neighborhood of infected vertices. Therefore, we can use a more efficient strategy
based on the previous algorithm. This strategy requires
to keep and constantly update a list $\mathcal{P}$ with the positions of all infected vertices where changes will take place. The
list update is simple. The position of a new infected is added at the end of the list. When a infected vertex becomes
susceptible, the last entry of the list is moved to the index of the cured vertex.

The total rate that a infected vertex becomes susceptible in the whole network is $R = \mu N_i$, where $N_i$ is the number of infected vertices.
Analogously, the total rate that one susceptible vertex is infected is given by $J = \lambda N_e$, where $N_e$ is the number 
of vertices emanating from infected nodes. So, the SIS dynamics can be simulated according to the 
algorithm proposed by Ferreira {\it et al.}~\cite{Ferreira12} as follows:
the step is incremented by $\Delta t =1/(R + J)$. With probability $p = R/(R + J)$ an infected vertex $i$ is selected 
randomly and turns it to susceptible. With complementary probability $q = J/(R + J)$ an infected vertex is selected at random 
and accepted with probability proportional to its degree. In the infection attempt, a neighbor of the selected vertex is randomly chosen and if 
susceptible, it is infected. Otherwise nothing happens and simulations run to the next
time step.

\subsection{Simulation of CP model}
\label{sec:CP_simulation}

The CP dynamics can also be efficiently simulated if a list of occupied vertex $\mathcal{P}$ is used analogously to the SIS algorithm.
The total rate of cure is also given by $R = \mu N_i$.
The total creation rate is $J = \lambda N_i$ \cite{Marrobook}. An infected vertex
$i$ is selected with equal chance. With probability $p = R/(J + R) = \mu/(\mu + \lambda)$ it is cured. With probability 
$q = J/(J + R) = \lambda/(\mu + \lambda)$ one of the $k_i$ neighbors of $i$ is selected and, if susceptible, is infected. 
The time is incremented by $\Delta t = 1/[N_i(\lambda + \mu)].$
Notice that infected vertices are selected independently of their degrees and with probability $n_i/k_i$ reach an already infected neighbor.

Although equivalent for strictly
homogeneous graphs ($k_i\equiv k~\forall~i$), the SIS and the CP models are very different for
heterogeneous substrates. The universality class
of CP and SIS is the same in homogeneous lattices. Both models
belong to the directed percolation universality class~\cite{henkel2008non}. Nevertheless, in complex networks, heterogeneity 
affects both models and, at the heterogeneous mean-field approach, they have different critical exponents 
(see discussion in Ref.~\cite{sander_phase_2013}).

In both SIS and CP models, the control parameter of the dynamics is the infection rate $\lambda$. 
In the thermodynamic limit, above a critical value $\lambda_c$ (epidemic threshold), a finite fraction of the population is infected. 
However, for $\lambda < \lambda_c$, the epidemic can not survive and the dynamics goes to an absorbing state where everyone is susceptible. 
Nevertheless, for a finite system the unique achievable stationary state is the absorbing state, even above the critical point, 
because of dynamical fluctuations. Some simulation methods were proposed in the literature to solve the issue of study the active
state in finite systems , as we will see in 
the next section. 

\subsection{Simulation of dynamical process with an absorbing state}
\label{sec:absorbing}

Mean-field approaches and field theory renormalization are the key analytical tools to 
investigate dynamical processes with absorbing-state phase transition~\cite{tauber2014}. 
While the former is only valid above the
upper critical dimension, application of the latter in physical
dimensions is hindered by large technical difficulties. For this reason,
most of our knowledge about the properties of absorbing-state phase
transitions is based in the computer simulation of different
representative models. 

The numerical analysis of these computer data
also represents a challenge mainly because of
finite size effects. In finite systems, any realization of the dynamics
reaches the absorbing state sooner or later, even above the
critical point, due to dynamic fluctuations. This difficulty was traditionally
overcome by starting with a finite initial density of active sites and
averaging only over surviving samples, {\it i.e.}, realizations which have not
yet fallen into the absorbing state~\cite{Marrobook,cardy88}. Analyzing the
quasistationary state defined by surviving averages, the
critical point and various critical exponents can be performed by
studying the decay of the survival average of different observables as a
function of the system size. However averaging over surviving runs is
computationally so inefficient since surviving configurations are increasingly rare at long times.

A much more efficient strategy is
provided by the quasistationary (QS) method, proposed by de Oliveira
and Dickman~\cite{DeOliveira05,PhysRevE.73.036131}, in which every time
the system  is on the verge of fall in an absorbing state, it jumps to an active
configuration previously stored during the simulation. This can be computationally implemented
by saving, and constantly updating a sample of the states already visited. 
The update is done by periodically replacing one of these configurations by the current one. 

The characteristic relaxation time is always short for epidemics on random 
networks due to the very small average shortest path~\cite{Newman10}.
The averaging time, on the other hand, must be large enough to 
guaranty that epidemics over the whole network was suitably averaged. It means 
that very long times are required for very low QS density (sub-critical phase)
whereas relatively short times are sufficient for high 
density states~\cite{DeOliveira05,DickmanJPA}.

Both equilibrium and non-equilibrium critical phenomena are hallmarked by 
simultaneous diverging correlation length and time which microscopically reflect 
divergence of the spatial and temporal fluctuations~\cite{henkel08}, 
respectively. Even tough a diverging correlation length has little sense on 
complex networks due to the small-world property~\cite{watts98}, the diverging 
fluctuation concept is still applicable. We used different criteria to determine 
the thresholds, relied on the fluctuations or singularities of the order 
parameters.

The QS probability $\bar{P}_n$, that does not depend on the initial condition, is defined as the probability that the system has 
$n$ occupied vertices in the QS regime, is computed during the averaging time 
and basic QS quantities, as lifespan and density of infected vertices, are 
derived from $\bar{P}_n$. Indeed, we have that 
$\rho = \frac{1}{N} \sum_n \bar{P}_n$ and $\tau = 1/\bar{P}_1$~\cite{DeOliveira05}, where $\tau$ is the lifespan of the epidemic.

Thus, thresholds for finite 
networks can be determined using the modified susceptibility~\cite{Ferreira12}
\begin{equation}
 \chi\equiv\frac{\lrangle{n^2}-\lrangle{n}^2}{\lrangle{n}}=
\frac{N(\lrangle{\rho^2}-\lrangle{\rho}^2)}{\lrangle{\rho}},
\label{eq:chi}
\end{equation}
that does exhibit a pronounced divergence at the transition point for 
SIS~\cite{Ferreira12,mata2013pair,Lee2013}  and contact 
process~\cite{RonanEPJB,Mata14} models on networks. 
The choice of the alternative definition, Eq.~(\ref{eq:chi}), instead of the
standard susceptibility $\tilde{\chi}=N(\lrangle{\rho^2}-\lrangle{\rho^2})$~\cite{henkel2008non} is
due to the peculiarities of dynamical processes on  complex networks.

\begin{figure*}[hbt!]
 \centering
 \includegraphics[width=7cm]{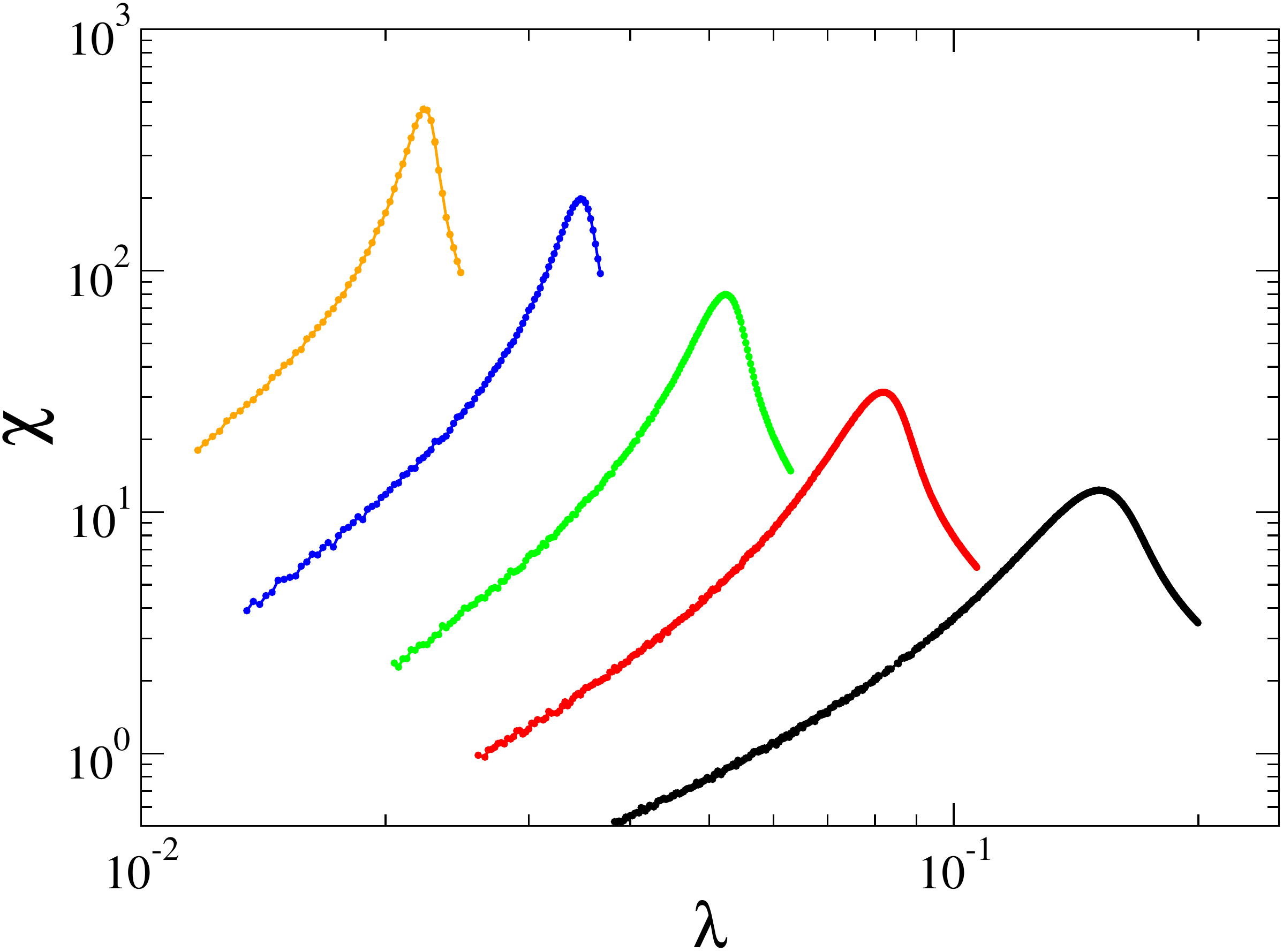}\quad
  \includegraphics[width=7cm]{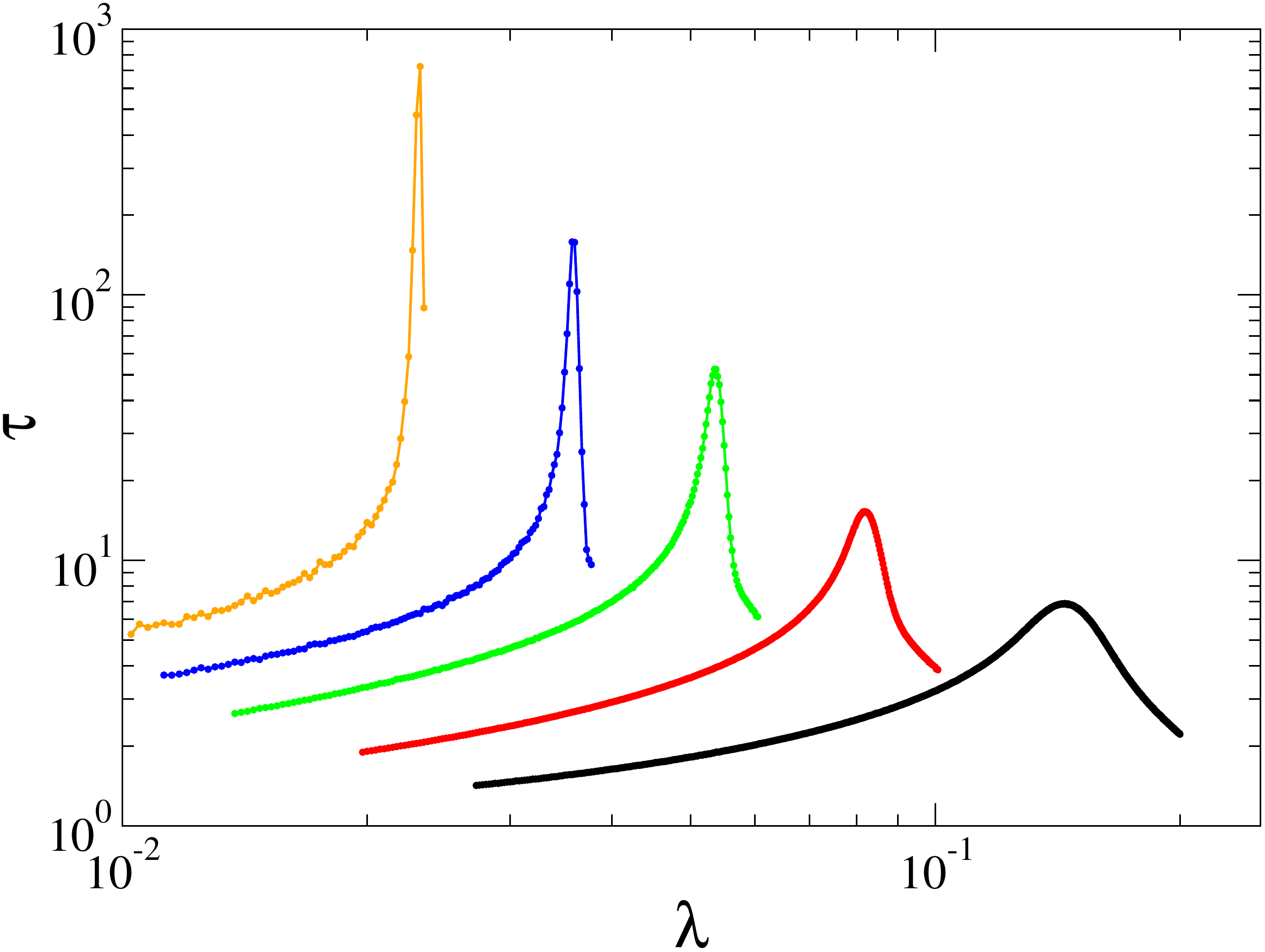}
  
 ~~~~~~~~ (a) ~~~~~~~~~~~~~~~~~~~~~~~~~~~~~~~~~~~~~~~~~~~~~~~~~~~~~~~~~~~~~ (b)
 \caption[Susceptibility and Lifespan against infection rate for SIS model on a single network for different sizes.]{(a)
 Susceptibility curves and (b) Lifespan curves against infection
rate for $N = 10^3, 10^4, 10^5, 10^6$ and $10^7$ (from the right to the left) used to determine
the thresholds in simulations (position of the peaks $\lambda_p$). 
 The network is constructed using the uncorrelated configuration model and has a power law degree distribution 
 $P(k) \sim k^{-\gamma}$ with
 $\gamma=2.75$ and minimum degree of connection equal to 3.}
 \label{fig:susc}
\end{figure*}

In a finite system of size $N$, $\chi$
shows a diverging peak at $\lambda=\lambda_p^{QS}(N)$, providing a
finite size approximation of the critical point. In the thermodynamic
limit, $\lambda_p^{QS}(N)$ approaches the true critical point with the
scaling form~\cite{binder2010monte}
\begin{equation}
  \lambda_p^{QS}(N) = \lambda_c + A_{QS} N^{-1/\nu},
  \label{eq:lambda_chapter1}
\end{equation}
as we can see for the SIS model in Figure \ref{fig:susc}(a).  The network was 
generated with the uncorrelated configuration model~\cite{Catanzaro05}, where 
vertex degree is selected from a power-law distribution $P(k) \sim k^{-\gamma}$ with a lower bound 
$k_0=3$. 

In the context of epidemic modeling on complex networks
\cite{Pastor-Satorras:2014aa}, Bogu\~{n}\'{a} {\it et. al.}~\cite{boguna2013nature}
proposed another strategy which considers the lifespan of spreading simulations starting from
a single infected site as a tool to determine the position of the
critical point. Each realization of the dynamical process is characterized by its
lifespan and its coverage $C$, where latter is defined as the fraction
of different sites which have been occupied at least once during the
realization. In the thermodynamic limit realizations can have either
\textit{finite} or \textit{infinite}, according to whether they proceed
below or above the critical point.  Endemic realizations have an
infinite lifetime and their coverage is equal to 1.  Finite realizations
have instead a finite lifetime and a coverage vanishingly small in the
limit of diverging size.

In finite systems this distinction is blurred, since any realization is
bound to end, reaching the absorbing state, although this can occur over long temporal scales.  In practice, a
realization is assumed as active whenever its coverage reaches a predefined
threshold value\footnote{The method is robust with respect to the coverage threshold (see Refs.\cite{boguna2013natureSupp,lifespan}).} $\Ct$, which was generally takes equal to 
$\Ct=0.5$.
Realizations ending before value $C=\Ct$ is reached are considered to be
finite. 

\begin{figure}[hbt!]
 \centering
 \includegraphics[width=7cm]{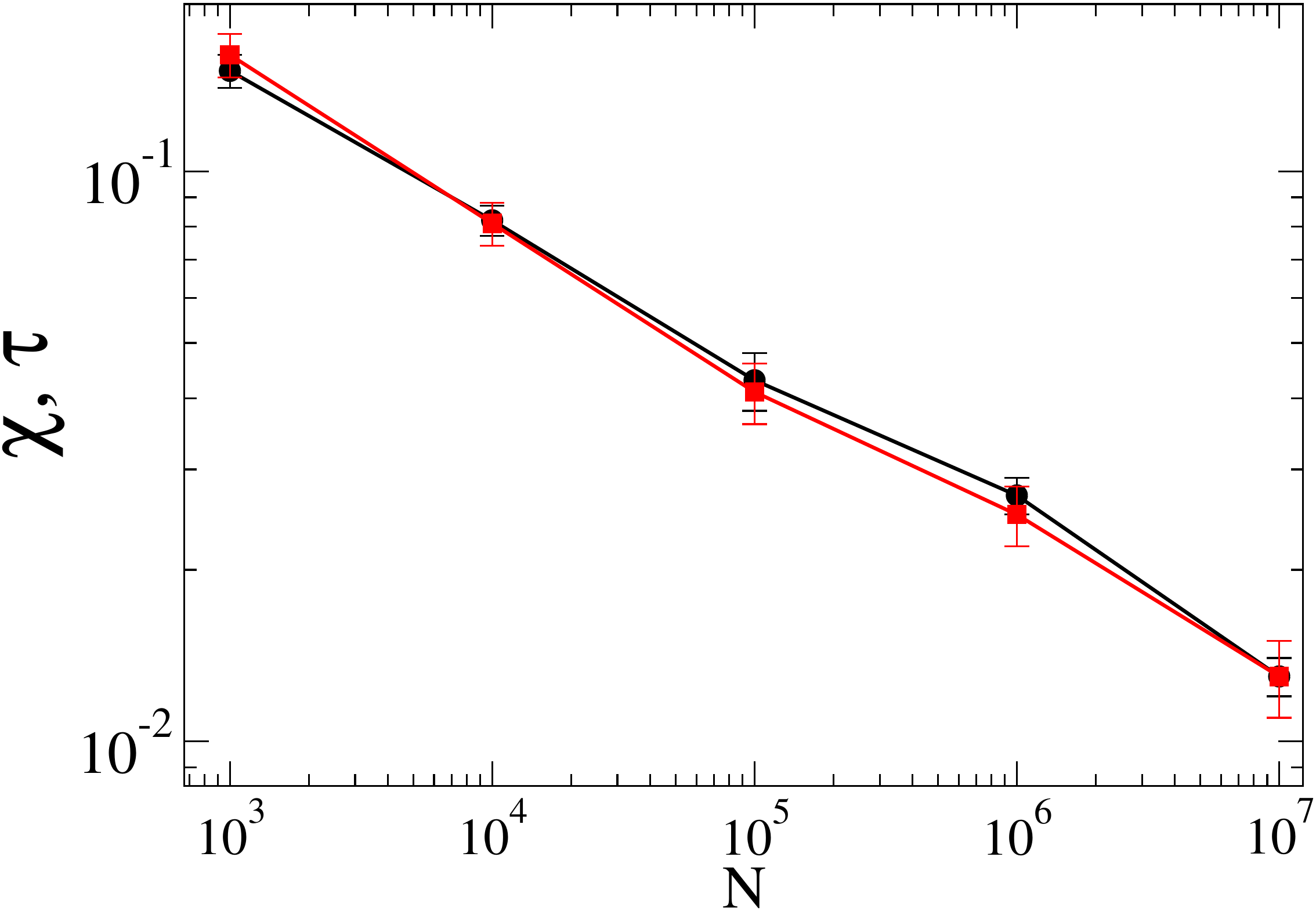}
 \caption[Critical Point]{The critical points $\lambda_p$ - obtained from the peaks of the curves shown in figure \ref{fig:susc} - versus N for both methods (susceptibility in 
red and lifespan in black) that agree very well.}
 \label{fig:tau}
\end{figure}

In this method the role of the order parameter is played by the
probability $\mathrm{Prob}(\lambda-\lambda_c,N)$ that a run is long-term,
while the role of susceptibility is played by the average lifetime of
finite realizations $\av{\tau}$.  For small values of $\lambda$ all
realizations are finite and have a very short duration $\tau$. As
$\lambda$ grows the average duration of finite realizations increases,
but for very large $\lambda$ almost all realizations are long-term, only
very short realizations remaining finite.  For this reason $\av{\tau}$
exhibits a peak for a value $\lambda_p(N)$ depending on $N$ and
converging to $\lambda_c$ in the thermodynamic limit. We can then use the average
lifespan to determine numerically the critical point, as shown in Figure \ref{fig:susc}(b).  
As we observe in Figure \ref{fig:tau}, the critical points as a function of network size
obtained by both methods are in very well agreement.

In reference~\cite{sander2016_2}, Sander and collaborators sumarize alternative methods for work around the problem 
related to the absorbing phase in simulations. They mentioned the reflecting boundary 
condition~\cite{sandpiles} that consists basically in avoiding the absorbing state by reverting the system to the
configuration that it was immediately before visit the absorbing state. Other strategy is to use a uniform external field that 
creates particles spontaneously at a given rate which disappears in the thermodynamic limit~\cite{Pruessner}. 
In addition, one can use the hub reactivation method on heterogeneous networks. If the system reaches the absorbing state, 
the dynamics starts again with the most connected vertex of the network infected. 

\section{Finite Size Scaling for CP model on heterogeneous networks}
\label{sec:HMF_CP_openproblem}

In Ref.~\cite{Castellano:2006}, Castellano and Pastor-Satorras derived the HMF theory for the CP dynamic in the
limit of infinite network size. They obtained the following scaling
\begin{equation}
\bar{\rho} \sim (\lambda-\lambda_c)^{\beta},~~~~~\beta=\max\left[1,\frac{1}{\gamma-2}\right].
\end{equation}
At the transition point $\lambda = \lambda_c$,
\begin{equation}
 \rho \sim t^{-\delta}, ~~~~~ \delta=\beta.
\end{equation}
Also the relaxation time scales as
\begin{equation}
 \tau \sim (\lambda-\lambda_c)^{-\nu_\parallel}, ~~~~~ \nu_\parallel=1.
\end{equation}
These exponents are also obtained using a pair HMF approximation as shown in reference~\cite{Mata14}.

It is not possible to check this predictions with numerical analysis because of the finite-size effects. A comparison
became possible using the finite-size scaling
(FSS) ansatz~\cite{cardy88}, adapted to the network topology, and they previously concluded that
CP dynamics on networks was
not described by the HMF approximation. However, it was assumed in Ref.~\cite{Castellano:2006} that
heterogeneous networks follow the same FSS 
known for regular lattices~\cite{Marrobook}. Indeed, the
FSS on networks is more
complicated than previously assumed. The behavior of the CP on networks of
finite size depends not only on the number of vertices $N$ but
also on the moments of the degree distribution~\cite{Castellano08}. This 
implies that, for scale-free networks, the scaling around the critical point
depends explicitly on how the largest degree $k_c$ diverges with the system size $N$.
Such dependence introduces very strong corrections to scaling.
However, when such corrections are suitably taken into account,
they showed that the CP on heterogeneous networks agrees
with the predictions of HMF theory with good accuracy~\cite{cpannealed}.

Ferreira and collaborators~\cite{FFCR11} started from Eq.~\eqref{eq:rhok1_cap2_CP}, and they considered, in addition, 
uncorrelated networks with $P(k|k') = kP(k)/k$, to obtain the 
equation for the overall density $\rho=\sum_k \rho_kP(k)$:
\begin{equation}
\label{eq:rhok1_cap2_CP1}
 \frac{d\rho(t)}{dt} = \rho(t) + \lambda \rho(t) \left[1 - \langle k \rangle^{-1}\sum_k kP(k)\rho_k(t)\right].
\end{equation}
A mean field theory for the  FSS can be obtained using the strategy proposed by Castellano and 
Pastor-Satorras~\cite{Castellano08}, in which the motion equation is mapped in a one-step process,
in the limit of very low densities, with transition rates
\begin{equation}
\label{eq:rates}
 \begin{array}{lll}
  W(n-1,n) & = & n \\
  W(n+1,n) & = & \lambda n \left[1 - \langle k \rangle^{-1}\sum_k kP(k)\rho_k(t)\right],
 \end{array}
\end{equation}
where $W(n,m)$ represents the transitions from a state with $m$ infected vertices to another state with $n$ infected vertices. 
In the stationary state, $d\rho(t)/dt = 0$, the Eq.~\eqref{eq:rhok1_cap2_CP1} read as~\cite{cpannealed}
\begin{equation}\label{eq:7_Cap1}
 \bar{\rho_k} = \frac{\lambda k \bar{\rho}/ \langle k \rangle}{1 + \lambda k \bar{\rho}/ \langle k \rangle}.
\end{equation}
Close to the criticality, when the density at long times is
sufficiently small such that $\bar{\rho}k_c \ll 1$,
Eq.~\eqref{eq:7_Cap1} becomes $ \bar{\rho_k} \simeq \lambda k \bar{\rho}/ \langle k \rangle$.
Substituting this result in Eq. \eqref{eq:rates}, one finds that the first-order approximation for the one-step processes is
\begin{equation}
\label{eq:rates2}
 \begin{array}{lll}
  W(n-1,n) & = & n \\
  W(n+1,n) & = & \lambda n (1 - \lambda g n/N),
 \end{array}
\end{equation}
 where $g = \langle k^2 \rangle /\langle k \rangle^2$. 
 
The master equation for a standard one-step process is \cite{vankampen}
\begin{equation}
  \label{eq:3chap1}
  \frac{d P_n}{dt} = \sum_{m} W(n,m) P_m(t) - \sum_m W(m,n) P_n(t).
\end{equation}
Substituting the rates \eqref{eq:rates2}, we find 
\begin{equation}
  \label{eq:master1}
  \frac{d P_n}{dt} = (n+1)P_{n+1}+u_{n-1}P_{n-1}-(n+u_n)P_n
\end{equation}
with $u_n =\lambda n (1-{n g})$. 
Since the probability for the process not to end up in the absorbing state up to time $t$, named 
survival probability, is given by $P_s(t) = \sum_{n\geq1}P_n(t)$, we can define the quasistationary (QS) 
distribution $\bar{P}_n$ as ~\cite{Marrobook}
 \begin{equation}
 \label{eq:qsdistribution}
  \bar{P}_{n} = \lim_{t \longrightarrow \infty} \frac{P_{n}(t)}{P_{s}(t)}  ~~~~~  (n \geq 1),
 \end{equation}
with $\bar{P}_0 \equiv 0$ and normalized condition $\sum_{n\geq1}\bar{P}_n = 1$ (see more details in section \ref{sec:absorbing}).
The solutions of the equation \eqref{eq:master1} have already been exhaustively investigated,  
then we merely report the results of Ref.~\cite{cpannealed} where it was found that the critical QS
distribution for 
large systems has the following scaling form\footnote{This was shown in Ref~\cite{cpannealed} for annealed networks, 
but this can also be applied for quenched large systems \cite{FFCR11}. In annealed networks, the vertex degrees are fixed 
while the edges are completely rewired between
successive dynamics steps implying that dynamical correlations are
absent and HMF theory becomes an exact prescription in the thermodynamic limit~\cite{Boguna09}.} 
\begin{equation}
\label{eq:scal_func}
 \bar{P}_n= \frac{1}{\sqrt{N/g}}f\left(\frac{n}{\sqrt{N/g}}\right),
\end{equation}
where $f(x)$ is as scaling function with the following properties: $f(x)\sim \exp(-ax)$ for $x\ll 1$, where $a$ is 
constant, and $f(x)\sim\exp(-x^2/2)$ for $x\gg1$. The critical quasistationary density scale as 
\begin{equation}
\bar{\rho}\sim (gN)^{-1/2}.
\end{equation}
 Similarly, the characteristic time scales as 
\begin{equation}
 \tau\sim \left(\frac{N}{g}\right)^{1/2}.
\end{equation}
For a network with degree exponent $\gamma$ and a
cutoff scaling with the system size as $k_c \sim N^{1/\omega}$, where $\omega = \text{max}[2,\gamma-1]$ for uncorrelated networks with power law degree distribution~\cite{Catanzaro05}, the factor $g$ scales for asymptotically large systems as $g\sim k_c^{3-\gamma}$ for $\gamma<3$ and $g\sim\mbox{const.}$ for $\gamma>3$. The result is a
scaling law $\rho\sim N^{-\hat{\nu}}$ and $\tau\sim N^{\hat{\alpha}}$ where the exponents $\hat{\nu}$ and $\hat{\alpha}$ are given by
\begin{equation}
  \hat{\nu} = \frac{1}{2}+\max\left(\frac{3-\gamma}{2\omega},0\right),~~
  \hat{\alpha} = \frac{1}{2}-\max\left(\frac{\gamma-3}{2\omega},0\right).
  \label{eq:exps}
\end{equation}

In Ref.~\cite{FFCR11}, Ferreira {\it et al.} investigated the CP on heterogeneous networks 
with power-law degree distribution by performing quasistationary simulations,
and concluded that heterogeneous
mean-field theory correctly describes the critical behavior of the contact process on quenched
networks. However, some  important questions remained unanswered. The transition point
$\lambda_c = 1$ predicted by this theory does not capture the dependence on the degree
distribution observed in simulations.  Sub-leading corrections to the finite-size scaling, undetected by the one-vertex HMF
theory, are quantitatively relevant for the analysis of highly heterogeneous networks ($\gamma \rightarrow 2$), for which
deviations from the theoretical finite-size scaling exponents were reported \cite{FFCR11}.

The HMF theory assumes that the number of
connections of a vertex is the quantity relevant to determine its state and
neglects all dynamical correlations.  But in reference~\cite{Mata14}, the authors present a pair HMF approximation, 
the simplest way to explicitly consider dynamical correlations, for the CP on
heterogeneous networks. Despite they found the same critical exponents obtained in the simple HMF approximation, the
corrections of the finite-size scaling were better, supporting that degree based theory estimate correctly the scaling exponents
of the contact process on scale-free networks.

\section{Epidemic threshold for the SIS model}
\label{sec:QMFversusHMF}

As we saw in the previous sections, distinct theoretical approaches were devised for the SIS 
and CP
models to determine an epidemic threshold
$\lambda_c$ separating an absorbing, disease-free state from an
active phase~\cite{Castellano10,Cator12a, Goltsev12, Ferreira12,
boguna2013nature, Lee2013, Chakrabarti08,mata2013pair}. The quenched mean-field (QMF)
theory~\cite{Chakrabarti08} explicitly includes the entire structure of the
network through its adjacency matrix while the heterogeneous mean-field (HMF)
theory~\cite{Pastor01,Pastor01b} performs a coarse-graining of the network
grouping vertices accordingly their degree. However, for the SIS model, both theories 
predicts different thresholds. The HMF theory predicts a
vanishing threshold for the range $2< \gamma\le 3$ while a
finite threshold is expected for $\gamma>3$. Conversely, the QMF theory states a
threshold inversely proportional to the largest eigenvalue of the adjacency
matrix, implying that the threshold vanishes for any value of $\gamma$
~\cite{Castellano10}. Regardless, Goltsev \textit{et al}.~\cite{Goltsev12} 
proposed that QMF theory predicts the threshold for an endemic phase, in which a
finite fraction of the network is infected, only if the principal eigenvector of
adjacency matrix is delocalized. In the case of a localized principal
eigenvector, that usually happens for large random networks with $\gamma>3$~\cite{odor2014localization}, the
epidemic threshold is associated to the eigenvalue of the first delocalized
eigenvector. For $\gamma<3$, there exists a consensus for SIS thresholds: both
HMF and QMF are equivalent and accurate for $\gamma<2.5$ while QMF works better
for $2.5<\gamma<3$~\cite{Ferreira12,mata2013pair}.

Lee {\it et. al.}~\cite{Lee2013} proposed that for a range $\lambda_{c}^{QMF} < 
\lambda < \lambda_c$ with a nonzero $\lambda_c$, the hubs in a random network 
become infected generating epidemic activity in their neighborhoods. This activity has a characteristic
lifespan $\tau(k,\lambda)$ depending on the degree $k$ and the infection rate $\lambda$. 
On networks where almost all 
hubs are directly connected the activity can be spread among them if the
lifespan $\tau(k,\lambda)$ is large enough. Then, above $\lambda_{c}^{QMF}$, the network is able
to sustain an endemic state due to the mutual reinfection of connected 
hubs. 

However, when hubs are not directly connected, the reinfection mechanism does not work
and high-degree vertices produce independent active domains. 
These independent domains were classified as rare-regions,  in which activity can last for very long periods increasing
exponentially with the domain size~\cite{Noest}.  This means, usually we have two distinct states: $\lambda > \lambda_c$ 
corresponds to a supercritical phase where the system is globally active and $\lambda < \lambda_c$ corresponds to an 
absorbing inactive state. However, the SIS dynamics running on top of power-law networks presents a region in which $\lambda$
is smaller than the epidemic threshold - but greater than a certain value below which the epidemic actually ends - where 
the activity survives for very long times. This results in a slow dynamics known as Griffiths 
phase~\cite{GPoriginal,GP,JuhaszCP}. The sizes of these active domains increase for increasing $\lambda$  leading to 
the overlap among them and, finally, to an endemic phase for $\lambda > \lambda_c$. In the thermodynamic limit these 
regions vanish because they decreases as soon as the network size increases~\cite{odor2013spectral,odor2014localization,cota2016}. 

This anomalous behavior in the subcritical phase was also investigated in reference~\cite{multipletransitions}. 
The authors used extensive simulations to show that the SIS model running on the top of power-law networks 
with $\gamma>3$ can exhibit multiple peaks in the susceptibility curve that are associated with large gaps in 
the degree distribution among the few most highly connected nodes, which permits the
formation of these independent domains of activity. However, if the number of hubs is large,
as occurs for networks with $\gamma <3$, the domains are directly connected and the activation
of hubs implies in the activation of the whole network.

The arguments presented by the authors of reference~\cite{multipletransitions} are in agreement with the 
scnario investigated in refs.~\cite{Goltsev12, Lee2013} that leads to the conclusion that 
the threshold to an endemic phase is finite in random networks with a power law degree distribution for $\gamma>3$. 
Inspired in the appealing arguments of Lee {\it et al.}~\cite{Lee2013},
Bogu\~n\'a, Castellano and Pastor-Satorras~\cite{boguna2013nature} 
reconsidered the problem and proposed a semi-analytical approach taking into account a long-range reinfection 
mechanism and found a vanishing epidemic threshold for $\gamma>3$.

As reported by Lee {\it et. al.}~\cite{Lee2013}, when the hubs on a network are directly connected, the activity can be spread 
throughout the network even in the limit $\lambda \rightarrow 0$. However, when higher degrees nodes are distant from each other 
these hubs are able to sustain local active domains around them and only with a nonzero $\lambda_c$, 
the endemic state is reached. Nevertheless, Bogu\~{n}\'{a}, Castellano and Pastor-Satorras \cite{boguna2013nature} revisited
the problem taking into account long range dynamic correlations in a coarse-grained time scale.
As explained in Ref.~\cite{boguna2013nature}, their approach states that a directed connection between hubs is not a 
necessary condition for reinfected them. There is a possibility of
``long-range'' reinfection since the network has a small-world property. 
In this approach, it was concluded that the epidemic threshold vanishes for random networks with 
$P(k)$ decaying slower than exponentially, in particular, $P(k) \sim k^{-\gamma}$ with any $\gamma$.
It was rigorously proved by Chatterjee and Durret~\cite{Chatterjee09} in the thermodynamic limit, for 
networks with degree distribution $P(k) \sim k^{-\gamma}$ with $\gamma > 3$. 

Afterward, Mountford and collaborators~\cite{mountford2013} expanded the result found in 
reference~\cite{Chatterjee09} including the range $2 < \gamma \leq 3$ of the degree exponent.  
They also analysed the behavior of the density of infected nodes in function of $\lambda$ close to the epidemic threshold
and they predicted analytical exponents which were also found in the numerical results of reference~\cite{spectralproperties}.

Recently, Castellano and Pastor-Satorras~\cite{Castellano2020} enlightened the understanding of the SIS dynamics elaborated 
a mathematical formulation of the mutual reinfection process, using the cumulative merging percolation (CMP) process
proposed by Menard and Singh~\cite{Menard2016}.  In this process, each node can be considered active with a certain 
probability. Inactive nodes play just as a bridge between active ones. A initial cluster with size 1 contains only
an active node but, in a interactive process, two clusters can colapse into one if the criterion of topological 
distance between them is satisfied. So the CMP process creates a cluster composed by a set of active nodes that were 
aggregated due to iteration of merging events.  Such nodes are part of the same connected component of the underlying 
network. 

The insight of Castellano and Pastor-Satorras~\cite{Castellano2020} were classified the hubs able to sustain the epidemic 
as these active nodes of the CMP process. Therefore they could relate this process to the reactivation of hubs that are not
directly connected. Consequently, they observed that the presence of a CMP giant component is related to an endemic 
stationary state. In their paper, they showed that the epidemic threshold does not behavior as QMF prediction but it 
vanished more slowly, with an exponent that decreases as soon as $\gamma$ increases. The dependence of the epidemic 
threshold with the network size that they found is in agreement with the asymptotic scaling found analytically by
Mountford and  collaborators~\cite{mountford2013} and recently by Huang and Durret~\cite{Huang2020}.

~\\

\section{Final remarks}
\label{sec:conclusions}

In this paper, we have reviewed the main features of the SIS and CP models, which have been very used to describe epidemic 
dynamics on complex networks.  Throughout this review we described both models, presented  distinct theoretical approaches 
devised for them, their advantages and disadvantages, and the main differences among them. We also exposed simulation 
techniques to analyze both models numerically. Since both models are examples to investigate absorbing phase transitions 
in complex networks, we presented the main simulation strategies to overcome the difficulty to study the active state of 
finite networks. Finally we reported the central issue for each model and we summarized the difficulties and discussions 
that came up in the literature related to these issues.  For SIS model, problems related to determine the epidemic threshold
on heterogeneous networks. For CP model, concerns related to the critical exponents and the degree distribution of the network.

Although there are some books and articles reviewing such models, the main idea of this manuscript 
is to provide, in summary, an overview of the main points of this subject: the theories and simulation techniques, 
as also the main concerns about the investigation of epidemic models with phase transitions to absorbing states running on top of complex networks.

The progress in this area grows incredibly fast and it is not possible to discuss
all recent results. But we try to mention just a few of many research lines. 
In the last decades, epidemic models
have also been studied in hypergraphs~\cite{hypergraphs}, temporal networks~\cite{temporal1,temporal2,temporal3}, 
metapopulations~\cite{metapopulation1,metapopulation2}
and also multiplex subtrates~\cite{multiplex1,multiplex2}. They analyzed, among other issues, 
epidemic spreading with awareness, social contagion, measures of epidemic control and how patterns of mobility affects the 
transmission of the disease. There are also studies about the impact of infectious period or recovery rates on epidemic 
spreading~\cite{impact1,impact2,impact3}, spectral properties of epidemic in correlated networks~\cite{spectralproperties}, 
the speed of disease spreading~\cite{speed}. It is also important to mention the challenges in modelling spreading diseases 
related to public health, global transmission~\cite{public_health,global_transmission} 
as well livestock and vector-borne diseases~\cite{livestock,vector_borne}. 
The relevance of studying epidemic models is also evident when faced with alarming situations such as the 
recent pandemic of COVID-19 caused by the new coronavirus~\cite{coronavirus1,coronavirus2,coronavirus3}.
These references and the others cited throughout the manuscript provide accurate 
studies for readers who wish to go deep into the subject.

\section*{Acknowledgments}
	This work was partially supported by CAPES, FAPEMIG and CPNq. 
	The author thanks the financial support from FAPEMIG (Grant No. APQ-02482-18) and CNPq 
	(Grant No. 423185/ 2018-7). The author also wishes to express her deepest gratitude for Silvio C. Ferreira 
	for reading this overview and providing relevant suggestions. 
\bibliographystyle{apsrev}
\bibliography{references}
\end{document}